\documentclass[
    amsmath,
    amssymb,
    amsfonts,
    reprint,
    aps,
    superscriptaddress,
    longbibliography,
    nofootinbib,
    floatfix,
]{revtex4-1}

\usepackage[utf8]{inputenc}
\usepackage[T1]{fontenc}
\usepackage{physics}
\usepackage{nicefrac}
\usepackage{bbm, bm}
\usepackage{makecell}
\usepackage{multirow}
\usepackage{booktabs}
\usepackage{overpic}

\usepackage[dvipsnames]{xcolor}

\usepackage[breaklinks=true, colorlinks]{hyperref}
\hypersetup{allcolors=blue}
\usepackage{cleveref}
\usepackage{orcidlink}

\newcommand{\GG}{{\mathbf{G}}}
\newcommand{\xx}{{\vb{x}}}
\newcommand{\kk}{{\vb{k}}}
\newcommand{\hh}{{\vb{h}}}
\newcommand{\rr}{{\vb{r}}}

\DeclareMathOperator*{\expect}{\mathbb{E}}
\DeclareMathOperator*{\Var}{{\text{Var}}}

\begin{document}

\title{Adiabatic Transport of Neural Network Quantum States}

\author{Matija Medvidović}
\email[Contact author: ]{mmedvidovic@ethz.ch}
\affiliation{Institute for Theoretical Physics, ETH Zürich, CH-8093 Zürich, Switzerland}

\author{Alev Orfi}
\affiliation{Center for Computational Quantum Physics, Flatiron Institute, 162 5th Avenue, New York, New York 10010, USA}
\affiliation{Center for Quantum Phenomena, Department of Physics, New York University, 726 Broadway, New York, New York 10003, USA}

\author{Juan Carrasquilla}
\affiliation{Institute for Theoretical Physics, ETH Zürich, CH-8093 Zürich, Switzerland}

\author{Dries Sels}
\affiliation{Center for Computational Quantum Physics, Flatiron Institute, 162 5th Avenue, New York, New York 10010, USA}
\affiliation{Department of Physics, Boston University, 590 Commonwealth Ave., Boston, Massachusetts 02215, USA}

\date{September 18, 2026}

\begin{abstract}
    Variational methods have offered controllable and powerful tools for capturing many-body quantum physics for decades. The recent introduction of expressive neural network quantum states has enabled the accurate representation of a broad class of complex wavefunctions for many Hamiltonians of interest. We introduce a first-principles method for building neural network representations of many-body excited states by adiabatically continuing eigenstates of simple Hamiltonians into the strongly correlated regime. With controlled access to the full many-body gap, we obtain accurate estimates of critical exponents. Successive eigenstate estimates can be run entirely in parallel, enabling precise targeting of excited-state properties without reference to the rest of the spectrum, opening the door to large-scale numerical investigations of universal properties of entire phases of matter.
\end{abstract}

\maketitle

\section{Introduction}

Neural quantum states (NQSs) have emerged as efficient and accurate representations of quantum states in many-body systems in recent years. Coinciding with the rapid progress of artificial intelligence for classical learning tasks, the NQS community has adopted and expanded state-of-the-art neural network models and pipelines~\cite{carrasquillaHowUseNeural2021, carleoMachineLearningPhysical2019, medvidovicNeuralnetworkQuantumStates2024, langeArchitecturesApplicationsReview2024}. Variational Monte Carlo (VMC) calculations with NQSs inherit a key advantage from the broader quantum Monte Carlo family of methods---They are \textit{first principles} and do not require an independent dataset~\cite{dawidMachineLearningQuantum2025} to produce quantitative predictions of correlated many-body states.

Initially used as trial states for variational ground-state optimization for spin systems~\cite{carrasquillaMachineLearningPhases2017, carleoSolvingQuantumManybody2017}, they have quickly expanded to fermionic problems on lattices~\cite{luoBackflowTransformationsNeural2019, robledomorenoFermionicWaveFunctions2022, malyshevAutoregressiveNeuralQuantum2023,  guSolvingHubbardModel2025, chenNeuralNetworkAugmentedPfaffian2025} and directly in real space~\cite{pfauInitioSolutionManyelectron2020, hermannDeepneuralnetworkSolutionElectronic2020, glehnSelfAttentionAnsatzAbinitio2023, lovatoHiddennucleonsNeuralnetworkQuantum2022, pesciaNeuralnetworkQuantumStates2022, pesciaMessagepassingNeuralQuantum2024, smithUnifiedVariationalApproach2024, kimNeuralnetworkQuantumStates2024, fosterInitioFoundationModel2025}. Due to the expressive power of neural networks, NQSs have shown particular promise in capturing difficult volume-law states~\cite{sharirNeuralTensorContractions2022}, where tensor network approaches~\cite{whiteDensityMatrixFormulation1992, schollwockDensitymatrixRenormalizationGroup2011} may struggle~\cite{wuVariationalBenchmarksQuantum2024}. As a consequence, they are increasingly being used as a method of choice to capture nonequilibrium phenomena with a nontrivial sign structure, in calculations at finite temperature~\cite{liDeepVariationalFree2025, nysRealtimeQuantumDynamics2024} and for real-time dynamics~\cite{schmittSimulatingDynamicsCorrelated2025, schmittQuantumPhaseTransition2022, schmittQuantumManyBodyDynamics2020, gutierrezRealTimeEvolution2022, medvidovicVariationalQuantumDynamics2023, sinibaldiUnbiasingTimedependentVariational2023, nysAbinitioVariationalWave2024, walleManybodyDynamicsExplicitly2024, sinibaldiTimedependentNeuralGalerkin2025}.

Despite the rapid progress of NQS methods, access to many-body excited states is often obstructed by prohibitive computational scaling. Excited states contain key information for a range of downstream physical observables such as gaps, critical exponents, and quasiparticle excitation spectra. These states are also a requirement for efficient and accurate impurity solvers for Green's function algorithms like dynamical mean-field theory~\cite{georgesHubbardModelInfinite1992, georgesDynamicalMeanfieldTheory1996} or GW~\cite{aryasetiawanGWMethod1998, reiningGWApproximationContent2018, golzeGWCompendiumPractical2019} at scale. Targeting the $n$th excited state, traditional approaches rely on projecting out contributions from lower states, adding $\sim n^2$ penalty terms to the calculation, and sacrificing accuracy and computational efficiency. Consequently, these methods require the first $n-1$ states before studying the $n$th excited state. More recently, a generalized variational principle was proposed to capture the subspace spanned by the chosen number of variational states~\cite{pfauAccurateComputationQuantum2024, kahnVariationalSubspaceMethods2025, hendryGrassmannVariationalMonte2025}. This approach allows simultaneous access to all $n$ states, but requires optimizing them collectively, leading to substantially higher computational overhead.

We introduce a method for accessing ground and excited states, based on adiabatic transport~\cite{kolodrubetzGeometryNonadiabaticResponse2017, kimVariationalAdiabaticTransport2024} projected onto a high-dimensional NQS parameter manifold. Starting from exact solutions of simple classical Hamiltonians, states are systematically \textit{dressed} with interactions. Unlike fine-tuning strategies that adjust preoptimized networks for nearby parameters~\cite{rendeSimpleLinearAlgebra2024, hernandes2025adiabatic}, our approach directly updates the variational parameters under Hamiltonian changes using efficient Monte Carlo estimators, allowing excited states to be targeted. Crucially, this approach is \textit{parallel} in excited states, allowing for the calculation of higher states without reference to the rest of the spectrum. The resulting finite-size scaling of the gap recovers the expected critical behavior at the transition, yielding precise estimates of the critical exponents and universal gap ratios in the underlying conformal field theory (CFT). By confirming the conclusions of Ref.~\cite{schulerUniversalSignaturesQuantum2016} via direct computation, we demonstrate that adiabatic transport captures the nontrivial operator content of three-dimensional (3D) Ising and Ising* CFTs near the thermodynamic limit. Because the method provides NQS representations at neighboring Hamiltonian parameters, geometric probes can be easily evaluated, offering generic signatures of critical behavior.

\begin{figure}[t]
    \centering
    \includegraphics[width=\linewidth]{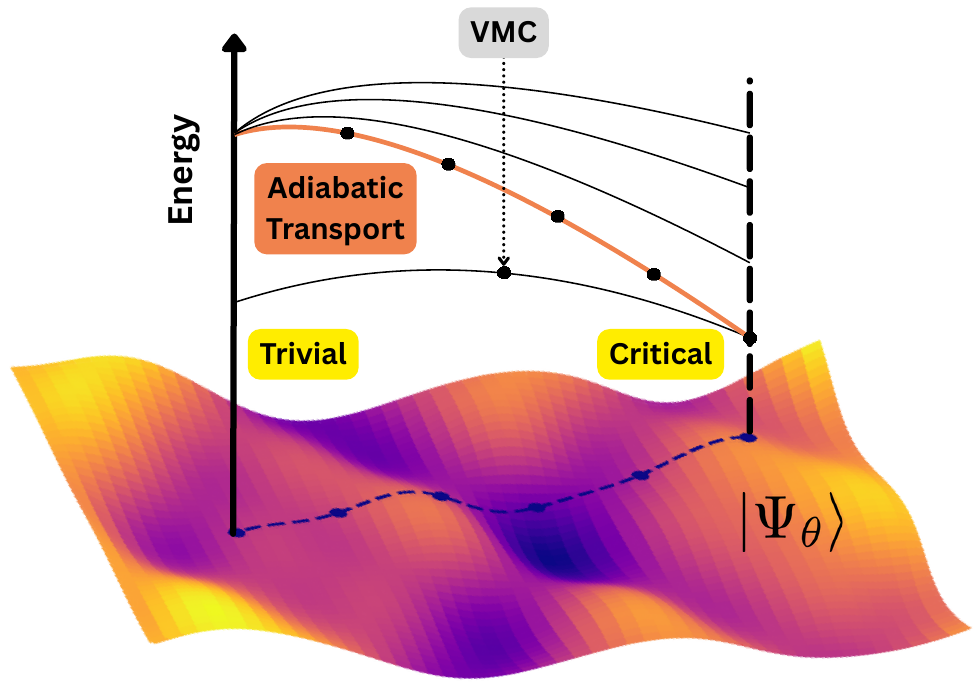}
    \caption{
        An overview of the adiabatic transport of an NQS ground state with traditional VMC ground-state optimization. Successive eigenstate approximations trace a curve in the space of parameters $\theta$ of the trial state $\Psi _\theta$.
    }
    \label{fig:diagram}
\end{figure}

\section{Methods}

Consider a many-body quantum system with a Hilbert space $\mathcal{H}$ spanned by an arbitrary computational basis $\{\ket{\xx}\}$ and described by a Hamiltonian $H_\lambda$ depending on a real parameter $\lambda$. For small changes in the Hamiltonian parameter, $\lambda \mapsto \lambda + \delta \lambda$, the instantaneous eigenstates and corresponding energies vary smoothly, unless a phase transition is crossed in the thermodynamic limit. While recent work~\cite{kimVariationalAdiabaticTransport2024} developed adiabatic transport methods for matrix product states, here we introduce a procedure for transporting NQS eigenstates as $\lambda$ is varied, enabling adiabatic transport into critical regions of two-dimensional (2D) systems. Figure~\ref{fig:diagram} illustrates the difference between adiabatic transport of eigenstates and traditional variational ground-state search.

We choose $H_0$ such that its relevant eigenstates can be determined exactly. The $n$th exact eigenstate $\ket{n_0}$ of $H_0$ serves as the initial condition of the adiabatic transport into the critical region, defined by $H_\lambda \ket{n_\lambda} = E _{n, \lambda} \ket{n_\lambda}$. The eigenstate $\ket{n_\lambda}$ for $\lambda > 0$ is approximated by an NQS variational state $\ket*{\Psi _\theta}$ by assuming that the network parameters $\theta$ themselves depend on $\lambda$,
\begin{equation}
    \ket{n_\lambda} \approx \ket*{\Psi _{\theta (\lambda)}} \propto \sum _\xx \psi _{\theta (\lambda)} (\xx) \ket{\xx} \; ,
\end{equation}
for a fixed functional form of the unnormalized trial wavefunction $\psi _\theta$.

We introduce the \textit{inverse power iteration} (IPI) method as a solver for updating general variational representations of quantum states. Recently, a similar method has been used as an improved optimizer for ground-state problems in Ref.~\cite{armegioiuFunctionalNeuralWavefunction2025}, showing faster convergence than natural gradient descent~\cite{sorellaGreenFunctionMonte1998, amariNaturalGradientWorks1998} in certain cases. The IPI method uses an approximate target eigenvalue $\omega \approx E$ of a Hamiltonian $H$ to recover the corresponding eigenstate $\ket{\Psi}$ such that $H \ket{\Psi} = E \ket{\Psi}$. Sequential eigenstate estimates are then refined by the well-known shift-and-invert procedure,
\begin{equation}
\label{eq:shift-and-invert}
    \ket{\Psi '} \propto \left( H - \omega \right)^{-1} \ket{\Psi} \; ,
\end{equation}
ignoring the normalization factor. Iterating the update in Eq.~\eqref{eq:shift-and-invert} recovers the eigenstate with the energy $E$ closest to the estimate $\omega$.

In the case of adiabatic transport, we use the IPI solver to propagate an approximate energy $E_\lambda$ and the corresponding eigenstate $\ket*{\Psi _{\theta (\lambda)}}$ from some $\lambda$ to $\lambda + \delta \lambda$. We set the target energy $\omega = \omega _\lambda$ to the first-order perturbative estimate
\begin{equation}
    \omega _\lambda = E_\lambda + \delta \lambda \; \bra{\Psi _{\theta (\lambda)}} \dv{H_\lambda}{\lambda} \ket{\Psi _{\theta (\lambda)}} \; ,
\end{equation}
and iterate the IPI scheme until convergence. After sufficiently many iterations, the resulting state $\ket{\Psi'}$ in Eq.~\eqref{eq:shift-and-invert} is identified with $\ket{\Psi'} \rightarrow \ket{\Psi_{\theta + \delta \theta}}$.

By directly substituting $\ket{\Psi'} \rightarrow \ket{\Psi_{\theta + \delta \theta}}$ into Eq.~\eqref{eq:shift-and-invert}, the update $\delta \theta$ can be determined, corresponding to a single IPI step on the variational manifold, as derived in Appendix~\ref{app:ipi}. This procedure leads to a linear system, $\GG \, \delta \theta = - f$, where
\begin{equation}
\label{eq:param-update}
\begin{gathered}
    \GG _{\mu \nu} = 2 \Re \; \bra{\partial _\mu \Psi _\theta} \left( H_\lambda - \omega_\lambda \right) \ket{\partial _\nu \Psi _\theta} \; , \\
    f _\mu = 2 \Re \; \bra{\partial _\mu \Psi _\theta} H_\lambda \ket{\Psi _\theta} \; .
\end{gathered}
\end{equation}
Here, the parameters $\theta$ are indexed with Greek indices and $\partial _\mu = \nicefrac{\partial}{\partial \theta ^\mu}$. In practice, the variational parameters are updated as $\theta ' = \theta + \eta \; \delta \theta$ with an empirically tuned mixing parameter $\eta$ to stabilize the iterations and enforce the condition that each $\delta \theta$ is small.

\begin{figure*}[t]
    \centering
    \begin{overpic}[width=\linewidth, trim=0 0em 0 0, clip]{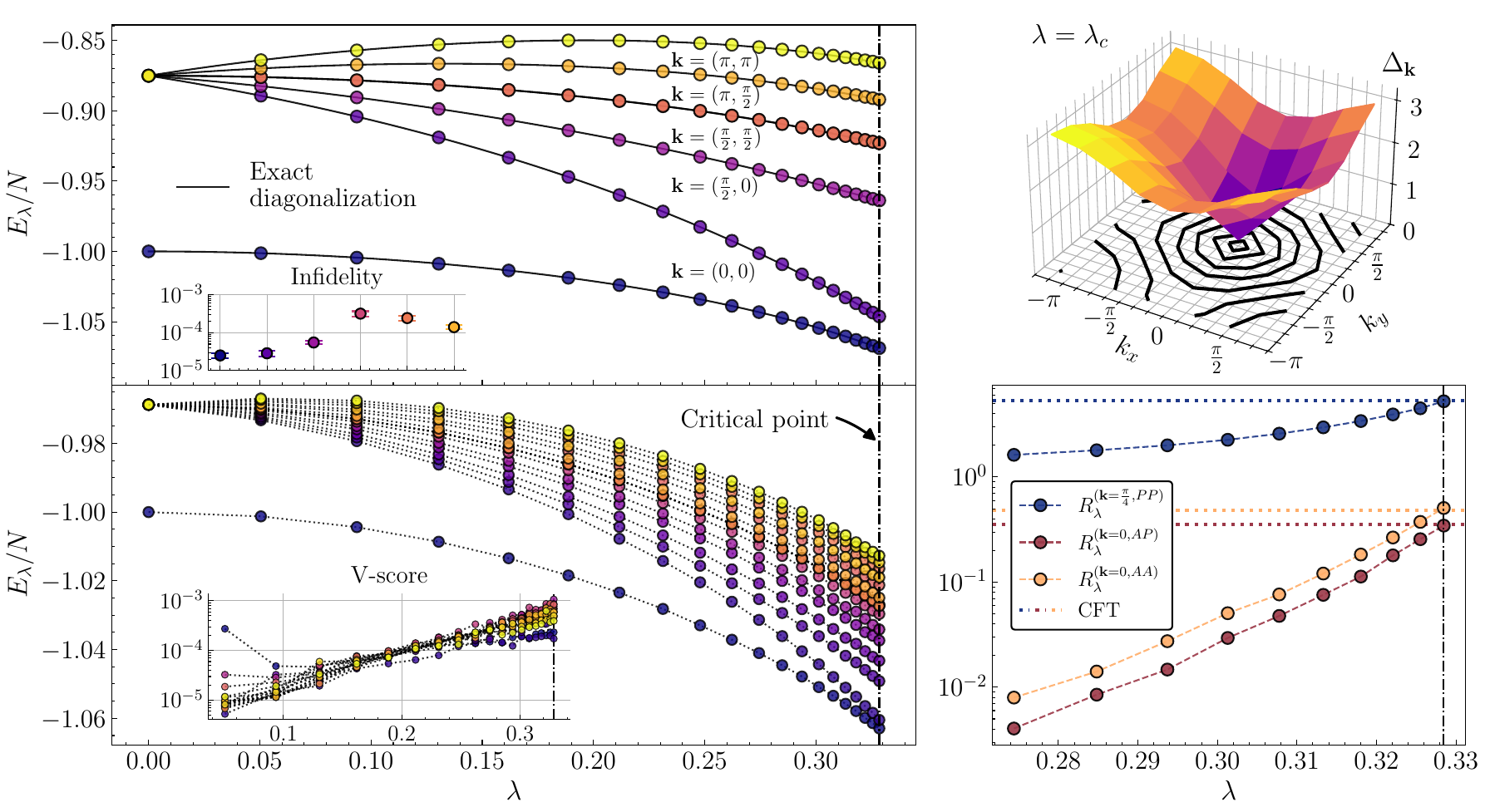}
        \put(9, 46){\fontsize{10}{12}\selectfont\textbf{(a)}}
        \put(88, 52){\fontsize{10}{12}\selectfont\textbf{(c)}}
        \put(9, 24){\fontsize{10}{12}\selectfont\textbf{(b)}}
        \put(93, 23){\fontsize{10}{12}\selectfont\textbf{(d)}}
    \end{overpic}
    \caption{
        Adiabatic transport of ground and excited states of the 2D TFIM to the critical point. Panel (a) shows the $4 \times 4$ lattice, with energies compared to exact diagonalization (solid lines) and an inset displaying the average infidelity for one representative eigenstate per degenerate manifold. Panel (b) shows the $8 \times 8$ lattice, with V-scores reported in the inset, and dispersion relation at the critical point shown in panel (c). Panel (d) shows the universal gap ratios $R_\lambda^X$ defined in Eq.~\eqref{eq:gap-ratio} for the $8\times 8$ 2D TFIM, in agreement with the CFT estimates of Ref.~\cite{schulerUniversalSignaturesQuantum2016} (horizontal lines) at the critical point $\lambda_c$.
    }
    \label{fig:higher-states}
\end{figure*}

\begin{figure*}[t]
    \centering
    \begin{overpic}[width=\linewidth, trim=0 0em 0 0, clip]{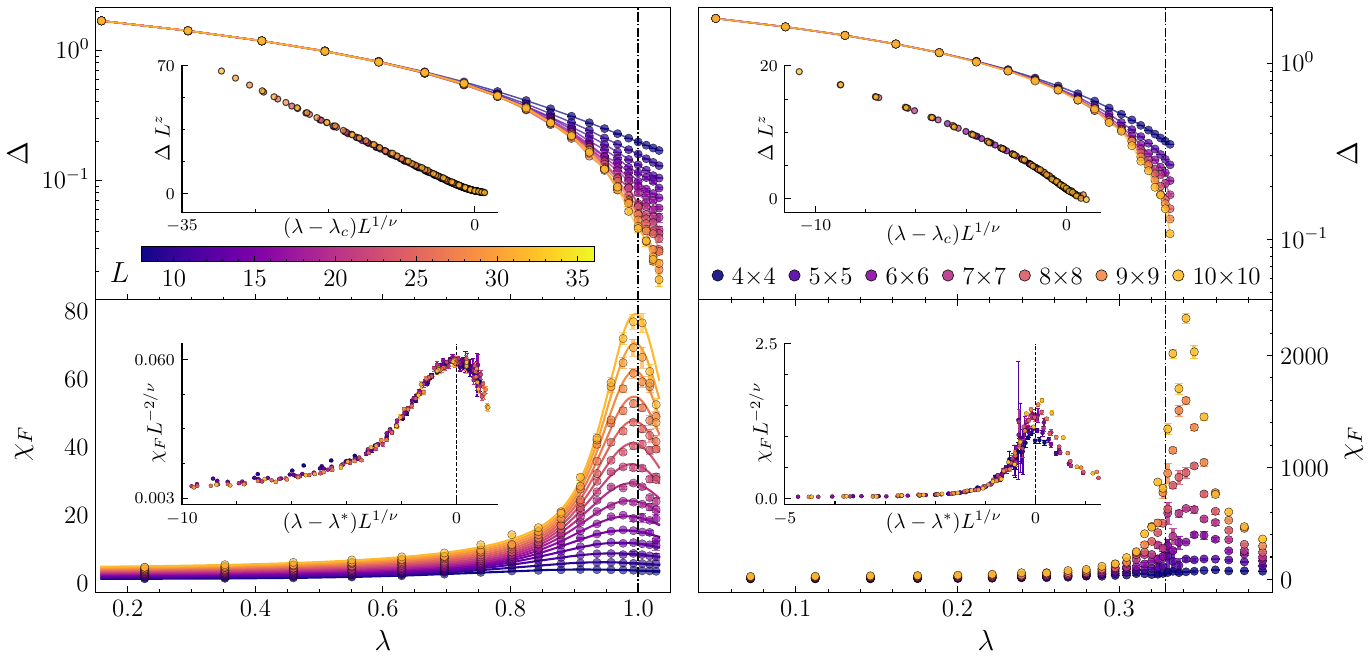}
        \put(42, 45.5){\fontsize{10}{12}\selectfont\textbf{(a)}}
        \put(88, 45.5){\fontsize{10}{12}\selectfont\textbf{(b)}}
        \put(8, 24){\fontsize{10}{12}\selectfont\textbf{(c)}}
        \put(51.5, 24){\fontsize{10}{12}\selectfont\textbf{(d)}}
    \end{overpic}
    \caption{
        Energy gap $\Delta$ vs $\lambda$ for 1D [panel (a)] and 2D [panel (b)] systems. The scaling $\Delta \sim L^{-z}$ at the critical point is used to extract the dynamical exponent $z$, while the insets display the finite-size data collapse used to determine the correlation-length exponent $\nu$. The extracted critical exponents are listed in Table~\ref{tab:exponents}. Panels (c) and (d) show the corresponding ground-state fidelity susceptibility for 1D and 2D systems, with the scaling collapse illustrated in the insets. In one dimension [panel (c)], the exact solution obtained via Jordan-Wigner transformation is shown as solid lines, demonstrating excellent agreement with the NQS results.
    }
    \label{fig:critical}
\end{figure*}

\section{Results}

The adiabatic transport method is validated on the prototypical transverse-field Ising model (TFIM) with periodic boundary conditions on a one-dimensional (1D) chain and a two-dimensional square lattice. The model Hamiltonian is
\begin{equation}
\label{eq:ising-hamiltonian}
    H _\lambda = -\lambda \sum_{\langle i, j \rangle} \sigma^z_i \sigma^z_j - \sum_i \sigma^x_i \; ,
\end{equation}
where $\sigma_i$ denotes Pauli operators acting on site $i$, and the first sum is taken over all nearest-neighbor pairs $i$ and $j$.

Defining the single-spin state $\ket{+}$ through $\sigma ^x \ket{+} = \ket{+}$, the ground state at $\lambda = 0$ takes the simple product form $\ket{0} = \mathop{\bigotimes} _i \ket{+} _i$ over all lattice sites. The single-particle excitation sector is spanned by momentum eigenstates $\ket{\kk}=\frac{1}{\sqrt{N}}\sum_i e^{i\kk\cdot\rr_i} \ket{i}$, where $\mathbf r_i$ is the position of site $i$ and $\ket{i}=\ket{-}_i\bigotimes_{j\neq i} \ket{+}_j$ denotes the configuration with a single spin flipped into the $\ket{-}$ state. These states provide exact low-lying excited states at $\lambda=0$ and are used as initial states for adiabatic transport. Transport of the ground and excited states is performed from a perturbatively small $\lambda_0$ to beyond the known critical points, $\lambda_c = 1$ in one dimension and $\lambda_c \approx 0.329$ in two dimensions \cite{bloteClusterMonteCarlo2002}.

The initial condition $\psi _0 (\xx)$ is built into the wavefunction as $\psi _{\theta(\lambda)}(\xx) = ( \psi _0 (\xx) ) ^a \times ( \phi _{\theta (\lambda)} (\xx) )^b$, where $\phi _{\theta (\lambda)} (\xx)$ is an NQS parametrized by $\theta(\lambda)$. Variational parameters $a$ and $b$ are optimized together with $\theta$, and $\psi _0 (\xx)$ is the exact amplitude at small $\lambda$. The NQS amplitude $\phi _{\theta (\lambda)} (\xx)$ is constructed as a custom architecture, combining convolutional~\cite{lecunDeepLearning2015} and residual~\cite{heDeepResidualLearning2015} layers with a top-level restricted Boltzmann machine (RBM)~\cite{salakhutdinovRestrictedBoltzmannMachines2007, carleoSolvingQuantumManybody2017, melkoRestrictedBoltzmannMachines2019}, maintaining the translational invariance of the wavefunction input $\xx$ introduced by periodic boundary conditions. We use a residual encoder $f_\alpha$ parametrized by $\alpha$ to generate hidden spins $\hh = f_\alpha (\xx)$ that are used to augment the existing input data as
\begin{equation}
    \ln \phi _\theta (\xx) = \texttt{RBM} _\beta \left( \left[ \xx, f_\alpha (\xx) \right] \right) \; ,
\end{equation}
where $[\cdots]$ is the concatenation operation and $\theta = \{ \alpha, \beta \}$. Further details of the network architecture and transport procedure are detailed in Appendix~\ref{app:net-details} and the Supplemental Material~\cite{supplement}.

Figure~\ref{fig:higher-states}(a) shows the adiabatic transport of the ground state and several excited states of the $4 \times 4$ TFIM to the critical point. The corresponding energies faithfully reproduce the results from exact diagonalization (ED) of the Hamiltonian using the Lanczos algorithm. Average infidelities over all $\lambda$ are shown in the inset, demonstrating the quality of the variational representation. For each degenerate manifold, one representative eigenstate is tracked, and its infidelity is evaluated with respect to the exact manifold. The excellent agreement with ED confirms that both ground- and excited-state wavefunctions are reliably captured across all values of $\lambda$.

We observe that energy crossings with states belonging to different discrete symmetry sectors do not destroy the transported state. In fact, energy levels shown in Fig.~\ref{fig:higher-states} undergo several crossings with odd-parity excited states that are not shown but are present in the full Hamiltonian spectrum. We also successfully transport states past the critical region in finite systems, indicating that with small enough steps, the method can resolve small gaps. If exact degeneracies appear during transport, parameter update equations become singular. However, we expect that using degenerate perturbation theory at the crossing should be sufficient to recover the transported state.

Far beyond the reach of ED, Fig.~\ref{fig:higher-states}(b) shows the transport of eigenstates of the $8 \times 8$ lattice. In the absence of ED reference data, we employ the V-score introduced in Ref.~\cite{wuVariationalBenchmarksQuantum2024} as a figure of merit. The V-score is a rescaled, dimensionless energy variance defined as $\text{V-score} = \flatfrac{N \Var H}{(E - E_\infty)^2}$, with $N$ being the total number of spins and $E_\infty = 0$ for spin systems such as the TFIM. We obtain reliable energies and wavefunctions for 15 eigenstates computed in parallel, with V-scores remaining below 0.001 even for the highest excited state at criticality. Despite being designed as a figure of merit for ground-state calculations, these V-score values suggest excellent agreement with the target state. Obtained values for higher excited states are comparable with state-of-the-art results for ground states of frustrated magnetic systems or fermionic lattice models.

Using the accurate ground and first excited states, we extract the critical exponents $z$ and $\nu$, which govern the scaling of the correlation length $\xi$ and the energy gap $\Delta$ near the critical point:
\begin{equation}
\label{eq:critical-exp-def}
    \xi \sim \left| \lambda - \lambda_c \right| ^{-\nu}
    \quad \text{and} \quad
    \Delta \sim \xi ^{-z}.
\end{equation}
Despite diverging at criticality in the thermodynamic limit, the correlation length is bounded by the system's linear size $L$ for finite systems. This leads to the finite-size scaling relation
\begin{equation}
\label{eq:gap-scaling}
    \Delta  = L^{-z} F \left( \left( \lambda - \lambda_c \right) L^{\nicefrac{1}{\nu}} \right) \; ,
\end{equation}
with a universal function $F$. The energy gap as a function of $\lambda$ is shown in Figs.~\ref{fig:critical}(a) and \ref{fig:critical}(b) for one and two dimensions, respectively. At the critical point, the scaling $\Delta \sim L^{-z}$ allows $z$ to be estimated via linear regression. Once $z$ is found, the exponent $\nu$ is obtained by collapsing the gap data across different system sizes according to Eq.~\eqref{eq:gap-scaling}, with the insets displaying the resulting high-quality collapse. Details of this procedure can be found in the Supplemental Material~\cite{supplement}. Numerical values of the critical exponents are given in Table~\ref{tab:exponents}, showing good agreement with the known values.

\begin{table}[t]
    \caption{
        Critical exponents $z$ and $\nu$ and CFT gap ratios from adiabatic transport. The critical exponents are compared with exact values calculated by conformal bootstrap~\cite{el-showkSolving3dIsing2014, changBootstrapping3dIsing2025, reehorstRigorousBoundsIrrelevant2022}, while the gap ratios are compared with the literature values of Ref.~\cite{schulerUniversalSignaturesQuantum2016}.
    }
    \label{tab:exponents}
    \centering
    \setlength{\tabcolsep}{5pt}
    \renewcommand{\arraystretch}{1.1}
    \begin{tabular}{c c c c}
        \toprule
        Dimension & Exponent & Exact & Transport \\
        \midrule
        \multirow{2}{*}{1D} & $z$   & 1       & 0.99(1)    \\
                            & $\nu$ & 1       & 1.024(3)   \\
        \midrule
        \multirow{2}{*}{2D} & $z$   & 1       & 1.03(2)    \\
                            & $\nu$ & 0.629\,97 & 0.6315(7)  \\
        \midrule
        \midrule
        Size & Ratio ($X$) & Literature & Transport \\
        \midrule
        \multirow{3}{*}{$8 \times 8$} & $(\mathbf k = \pi/4, \mathrm{PP}) $ & 5.305 & 5.26(3) \\
                                      & $(\mathbf k = 0, \mathrm{AP}) $   & 0.352 & 0.345(6) \\
                                      & $(\mathbf k = 0, \mathrm{AA}) $        & 0.484 & 0.507(7) \\
        \bottomrule
    \end{tabular}
\end{table}

The availability of NQS wavefunctions at different $\lambda$ enables direct estimation of the ground-state fidelity susceptibility, a general probe of quantum phase transitions that does not rely on prior knowledge of the order parameter \cite{wangFidelitySusceptibilityMade2015}. It is defined as
\begin{equation}
\label{eq:fid_sus}
    \chi _{\mathcal{F}} (\lambda) = - \lim_{\epsilon \to 0} \pdv[2]{\epsilon} \ln \mathcal{F}_n(\lambda, \lambda+\epsilon) \; ,
\end{equation}
where $\mathcal{F}(\lambda,\lambda+\epsilon) = \left| \braket{\Psi_n(\lambda)}{\Psi(\lambda + \epsilon)} \right|^2 $. In the thermodynamic limit, $\chi_{\mathcal{F}}$ diverges at criticality, following universal scaling laws \cite{polkovnikovUniversalAdiabaticDynamics2005, camposvenutiQuantumCriticalScaling2007}, and can be estimated from neighboring fidelities as $\chi _{\mathcal{F}} \approx -\frac{1}{\epsilon^2} \ln \left( \mathcal{F} (\lambda-\epsilon) \mathcal{F} (\lambda+\epsilon) \right)$. As shown in Figs.~\ref{fig:critical}(c) and \ref{fig:critical}(d), this estimate reproduces the expected divergence in both one and two dimensions, with the scaling collapse visible in the insets. In one dimension, the TFIM can be solved through a Jordan-Wigner transformation to a free fermion model \cite{sachdevQuantumPhaseTransitions2011}, and our results are in excellent agreement with the exact solution [solid lines in panel (c)]. See the Supplemental Material~\cite{supplement} for more details.

At criticality, the physics is governed by CFT. The spectra of $(2+1)$-dimensional CFTs on a spatial torus are expected to be universal \cite{schulerUniversalSignaturesQuantum2016}, but unlike their two-dimensional counterparts, they are not fixed by radial quantization and are analytically inaccessible \cite{difrancescoConformalFieldTheory1997}.

In Fig.~\ref{fig:higher-states}(d), we show that adiabatic transport reproduces universal torus gap ratios of the 3D Ising and Ising* CFTs. Using the lowest excitation with fully periodic boundary conditions as the reference gap, $\Delta_\lambda^{\mathrm{ref}}=E_\lambda^{(\mathbf{k}=0,\mathrm{PP})}-E_\lambda^{(\mathrm{GS},\mathrm{PP})}$, the gap ratio is defined as
\begin{equation}
\label{eq:gap-ratio}
    R_\lambda^X=\frac{E_\lambda^X -E_\lambda^{(\mathrm{GS},\mathrm{PP})}}{\Delta_\lambda^{\mathrm{ref}}}
    \; .
\end{equation}
Here, $X$ denotes either a finite-momentum excitation with periodic boundary conditions (PP) or the lowest-energy state obtained by imposing antiperiodic boundary conditions along one (AP) or both (AA) torus cycles. The former is associated with a conformal descendant, while AP and AA boundary conditions implement $\mathbb{Z}_2$ defect lines wrapping one or both cycles of the torus.

The estimates of the CFT gap ratios at the critical point for the $8\times 8$ results are presented in Table~\ref{tab:exponents}.
The finite-momentum and AP ratios agree well with the estimates of Ref.~\cite{schulerUniversalSignaturesQuantum2016}, while the AA ratio shows a larger deviation, consistent with stronger finite-size effects from two intersecting defect lines.
Together with the critical-exponent estimates, these results show that the variational NQS approach resolves both momentum and defect excitations near the transition.

\section{Conclusion}

We have introduced an accurate transport method to access NQS excited states, naturally extending the existing reliable VMC toolbox. Generalized stochastic parameter update equations based on the shift-and-invert procedure, used with simple perturbative estimates of target energies, enable the parallel preparation of NQS excited states. Without sacrificing accuracy, this feature eliminates the quadratic computational overhead introduced by enforcing orthogonality constraints. Benchmarks against the integrable (one-dimensional) and large nonintegrable (two-dimensional) TFIM reveal that adiabatic transport, coupled with NQS, can access both low-lying excited states and universal critical physics through accurate estimates of the many-body gap and the critical exponents.

On-demand access to excited states removes important roadblocks in precision many-body calculations. They are a key ingredient of quasiparticle excitation spectra on top of highly correlated states. The screening of candidate materials for a target property, such as the estimation of optical band gaps~\cite{zhaoVariationalExcitationsReal2019, huntQuantumMonteCarlo2018}, can benefit significantly from precise access to higher states. Similarly, finite-temperature and real-time solvers indirectly rely on excited states to push our understanding of thermal properties as well as prethermalization. Adiabatic transport itself has become a key computational benchmark for near-term practical quantum advantage~\cite{kingBeyondclassicalComputationQuantum2025}.

Having recovered universal CFT gap ratios for the Ising and Ising* sectors, adiabatic transport has shown promise as a numerical probe of universal physics. As topological states are characterized by robust ground-state degeneracies on the torus, NQS transport methods may provide a route to resolving small gaps in other topologically ordered systems. Exploiting the high expressive power of NQS to represent correlated \textit{spectra} is well positioned to be the next frontier of computationally driven physical insight. Adiabatic transport offers a universal and rapidly scalable computational framework to reach this goal. We are excited to see which challenging open problems it will attack next.

\section*{Acknowledgments}
M.M. acknowledges many useful discussions with Jannes Nys. A.O. acknowledges support from the Pierre Hohenberg Graduate Scholar Fellowship and computational resources provided by the Flatiron Institute. The Flatiron Institute is a division of the Simons Foundation. D.S. thanks AFOSR for support through Award No. FA9550-25-1-0067.

\section*{Data Availability}

All simulations were performed on graphical processing units using the JAX~\cite{bradburyJAXComposableTransformations2018} library for array manipulation and automatic differentiation. Equinox~\cite{kidgerEquinoxNeuralNetworks2021} was used for neural network design and Optax for optimization. Data were postprocessed using NumPy~\cite{harrisArrayProgrammingNumPy2020} and SciPy~\cite{virtanenSciPy10Fundamental2020}. The plots were produced using the Matplotlib~\cite{hunterMatplotlib2DGraphics2007} library. The code needed to reproduce the results in this work and explore new ones can be found in the repository in Refs.~\cite{code, codeGithub}.

\appendix

\section{Inverse power iteration}
\label{app:ipi}

In this Appendix, we give a more detailed treatment of the IPI. As noted in the main text, given an estimate of the target eigenvalue $\omega$, we produce a sequence of eigenstate estimates $\ket{\Psi _k}$ such that $\ket{\Psi _{k+1}} \propto (H - \omega)^{-1} \ket{\Psi _k}$. After sufficiently many iterations, this procedure extracts a target eigenstate $\ket{m}$ with energy $E_m$ if the starting state $\ket{\Psi _0} = \sum _n c_n \ket{n}$ has a nonzero overlap with it,
\begin{equation}
\begin{gathered}
    \lim _{k \rightarrow \infty} \ket{\Psi _k} = \lim _{k \rightarrow \infty} (H-\omega)^{-k} \ket{\Psi _0}\\
    = \lim _{k \rightarrow \infty}  \sum _n c_n (E_n - \omega)^{-k} \ket{n} \propto \ket{m} \; .
\end{gathered}
\end{equation}
To project one IPI onto the variational manifold, we require that parameter updates $\delta \theta$ satisfy
\begin{equation}
\label{eq:small-param-cond}
    \ket{\Psi _{\theta + \delta \theta}} \approx \ket{\Psi _\theta} + \sum _\mu \delta \theta ^\mu \ket{\partial _\mu \Psi _\theta} \propto (H - \omega)^{-1} \ket{\Psi _\theta} .
\end{equation}
We multiply Eq.~\eqref{eq:small-param-cond} by $(H - \omega)$ from the left and expand the solution in the tangent space basis $\ket{\partial _\mu \Psi _\theta}$, recovering
\begin{equation}
\begin{gathered}
    \sum _\nu \Re \; \left\{ \bra{\partial _\mu \Psi _\theta} \left( H - \omega \right) \ket{\partial _\nu \Psi _\theta} \right\} \delta \theta ^\nu \\
    = - \Re \; \left\{ \bra{\partial _\mu \Psi _\theta} H \ket{\Psi _\theta} \right\} \; ,
\end{gathered}
\end{equation}
after making use of $\partial _\mu \braket{\Psi _\theta} = 2 \Re \braket{\partial _\mu \Psi _\theta}{\Psi _\theta} = 0$ and the assumption that the parameters are real. The last equation leads to the parameter update rule $\delta \theta = -\GG^{-1} f$, with definitions in Eq.~\eqref{eq:param-update}. The matrix $\GG$ and the vector $f$ must be evaluated by Monte Carlo methods as
\begin{equation}
\label{eq:ipi-mc}
\begin{gathered}
    \GG _{\mu \nu} = 2 \, \Re
    \expect _{\xx \sim |\Psi _\theta |^2} \left[ \frac{\braket{\partial _\mu \Psi _\theta}{\xx}}{\braket{\Psi _\theta}{\xx}} \frac{\bra{\xx} \left( H - \omega \right) \ket{\partial _\nu \Psi _\theta}}{\braket{\xx}{\Psi _\theta}} \right] \; , \\
    f_\mu = 2 \Re
    \expect _{\xx \sim |\Psi _\theta |^2} \left[ \frac{\braket{\partial _\mu \Psi _\theta}{\xx}}{\braket{\Psi _\theta}{\xx}} \frac{\bra{\xx} H \ket{ \Psi _\theta}}{\braket{\xx}{\Psi _\theta}} \right] \; .
\end{gathered}
\end{equation}
We identify familiar ingredients of traditional VMC calculations in Eq.~\eqref{eq:ipi-mc} as the local energy $E_\text{loc} (\xx)$ and the NQS \textit{Jacobian} $\mathcal{J} _\mu (\xx)$, and define the \textit{projected Jacobian} $\mathcal{P}_\mu (\xx)$ as
\begin{equation}
\label{eq:ipi-matrices}
\begin{gathered}
    E_\text{loc} (\xx) = \frac{\bra{\xx} H \ket{ \Psi _\theta}}{\braket{\xx}{\Psi _\theta}}
    \; , \quad
    \mathcal{J} _\mu (\xx) = \frac{\braket{\xx}{\partial _\mu \Psi _\theta}}{\braket{\xx}{\Psi _\theta}}
    \quad \\ \text{and} \quad
    \mathcal{P}_\mu (\xx) = \frac{\bra{\xx} \left( H - \omega \right) \ket{\partial _\mu \Psi _\theta}}{\braket{\xx}{\Psi _\theta}} \; .
\end{gathered}
\end{equation}
All quantities defined in Eq.~\eqref{eq:ipi-matrices} have efficient Monte Carlo estimators provided that the Hamiltonian $H$ is sparse enough to have only polynomially many connected basis elements to $\ket{\xx}$, which is a requirement for ground-state VMC calculations as well. We refer readers interested in the full treatment of local energies and NQS Jacobians to Refs.~\cite{medvidovicNeuralnetworkQuantumStates2024, langeArchitecturesApplicationsReview2024, carrasquillaHowUseNeural2021, dawidMachineLearningQuantum2025}. After algebraic manipulation, we write the estimator for the projected Jacobian as
\begin{equation}
    \mathcal{P}_\mu (\xx) = \partial _\mu E_\text{loc} (\xx)  + \mathcal{J}_\mu (\xx) (E_\text{loc} (\xx) - \omega) \; .
\end{equation}
We note that all of the derivatives in the resulting expressions can be evaluated using efficient automatic differentiation. The intractable norm of the variational state cancels out in all expressions in Eqs.~\eqref{eq:ipi-matrices}~and~\eqref{eq:finite-sample-system}.

In the finite-sample approximation using Monte Carlo samples $\{ \xx_1, \ldots , \xx _{N_s} \}$, drawn so that $\expect _{\xx \sim |\Psi _\theta |^2} [A(\xx)] \approx \frac{1}{N_s} \sum _{i=1} ^{N_s} A(\xx_i)$ holds, the parameter update equation reduces to a simple linear system. Defining matrices
\begin{equation}
\label{eq:finite-sample-system}
    J_{i \mu} = \frac{\mathcal{J}_\mu (\xx _i)}{\sqrt{N_s}} \; , \quad
    P_{i \mu} = \frac{\mathcal{P}_\mu (\xx _i)}{\sqrt{N_s}} \quad \text{and} \quad
    \varepsilon _{i} = \frac{E _\text{loc} (\xx _i)}{\sqrt{N_s}} \; ,
\end{equation}
that linear system reads $J^\top P \; \delta \theta = - J^\top \varepsilon$.

For any given inverse power iteration, it is beneficial to employ the Woodbury identity~\cite{chenEmpoweringDeepNeural2024, rendeSimpleLinearAlgebra2024} whenever the number of variational parameters $P$ exceeds the number of Monte Carlo samples $N_s$,
\begin{equation}
\label{eq:param-update-inverse}
\begin{aligned}
    \delta \theta &= - (J^\top P + \gamma \mathbbm{1} _{P} ) ^{-1} J^\top \varepsilon \\
    &= - J^\top (P J^\top + \gamma \mathbbm{1} _{N_s} ) ^{-1} \varepsilon \; ,
\end{aligned}
\end{equation}
to invert larger matrices, with an optional diagonal shift $\gamma$ for numerical stability.

Although the choice of the linear solver is arbitrary in principle as long as the parameter update takes the form $\theta ' = \theta - \eta \; \GG ^{-1} f$, in practice, we found that a modified pseudoinverse offers the best balance of accuracy and speed. The solver we use has been developed in previous research by some of the authors in Ref.~\cite{medvidovicVariationalQuantumDynamics2023}, building on insights in Ref.~\cite{schmittQuantumManyBodyDynamics2020}. More details can be found in the Supplemental Material~\cite{supplement}.

Shift-and-invert is guaranteed to converge exponentially for dense matrices. In the case of stochastically estimated $J$ and $P$, we observe that convergence depends on the number of Monte Carlo samples and the conditioning of $J$ as well. Our tests and those of Ref.~\cite{armegioiuFunctionalNeuralWavefunction2025} suggest that the stochastic version inherits IPI convergence properties of the dense version under sufficient Monte Carlo samples. Determining exact conditions for convergence and more detailed measurements is left for future work.

\section{Neural network architecture}
\label{app:net-details}

The variational wavefunction is defined as the residual RBM~\cite{salakhutdinovRestrictedBoltzmannMachines2007, melkoRestrictedBoltzmannMachines2019, carleoSolvingQuantumManybody2017} multiplied by the initial condition wavefunction, as outlined in the main text. The architecture is based on a simple, shallow RBM trial state commonly used in VMC calculations. We use an adapted version of the RBM that is invariant with respect to translational lattice symmetries,
\begin{equation}
\label{eq:rbm-def}
    \textup{RBM} _{(w, b)} (\xx) = \sum _{k, \rr} \ln \cosh \left( b^k + (w^k \ast \xx ) _{\rr} \right) \; ,
\end{equation}
due to the appropriate convolution operation $\ast$ over lattice coordinates $\rr$. In Eq.~\eqref{eq:rbm-def}, biases $b$ and weights $w$ can be optimized to represent different states of lattice spins $\xx$. The RBM on its own is a parameter-efficient ansatz capable of capturing many interesting correlated states.

\begin{figure}[t]
    \centering
    \includegraphics[width=\linewidth]{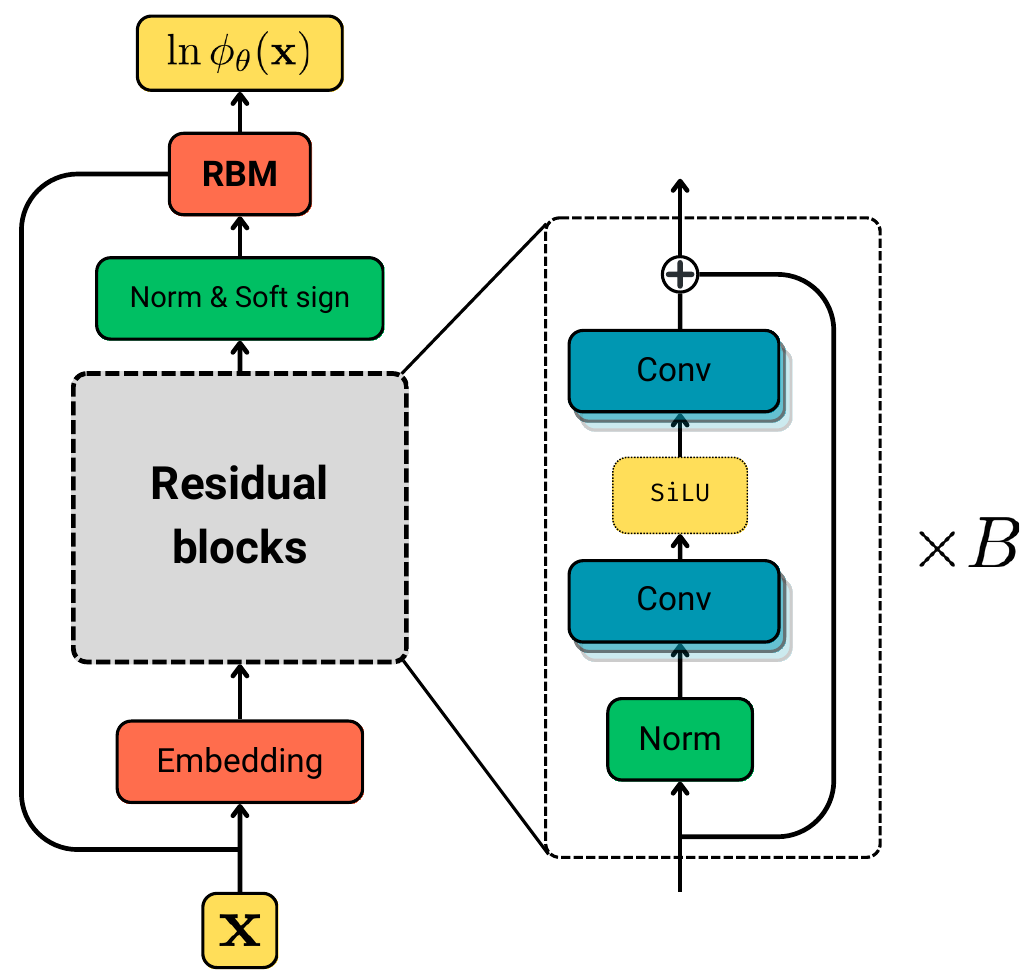}
    \caption{
        The sublayer internal connectivity of the residual RBM NQS architecture used in the adiabatic transport simulations. This diagram shows the NQS amplitude factor of the full wavefunction.
    }
    \label{fig:nn-diagram}
\end{figure}

However, we found that capturing excited states requires increased expressivity beyond RBMs. Therefore, we supplement the RBM architecture with a custom convolutional encoder~\cite{lecunDeepLearning2015} $f_\alpha$ parametrized by $\alpha$. The encoder outputs additional \textit{hidden} lattice spin configurations $\vb{y} = f_\alpha (\xx)$, which are stacked on top of the input configuration $\xx$. The stacked spin configurations are passed into an RBM with complex parameters, producing the final log-amplitude $\ln \psi _\theta (\xx)$. Augmenting the existing RBM architecture with hidden spins in this way is inspired by \textit{backflow} methods for fermionic Hamiltonians and allows us to make the overall model more expressive without sacrificing any of the components that make the RBM a useful inductive bias in the first place. Mathematically, we replace the RBM input $\xx$ in Eq.~\eqref{eq:rbm-def} with an augmented spin variable $\Tilde{\xx}$ with $d+1$ \textit{channels}
\begin{equation}
    \Tilde{\xx} = \left[ \, \xx ,\, f ^{(1)} _\alpha (\xx) ,\, \ldots ,\, f ^{(d)} _\alpha (\xx) \, \right] \; ,
\end{equation}
where $[\cdots]$ denotes the concatenation along a new \textit{feature} or \textit{channel} dimension. The internal connectivity of the resulting neural network is shown in Fig.~\ref{fig:nn-diagram}.

The rest of this Appendix is devoted to laying out the structure of the convolutional encoder $f_\alpha$. First, we lift the internal representation of each spin from $\sigma_i \in \{-1, +1\}$ to $\hh _i = \sigma _i \otimes \vb{u}$, where $\vb{u} \in \mathbbm{R}^d$ is a trainable embedding vector. The embedding vector is shared between all lattice sites to preserve any lattice symmetries.

After embedding, the lifted spin representation is passed through $B$ \textit{residual} blocks in sequence. Each block is a two-layer convolutional network with a \textup{SiLU}~\cite{elfwingSigmoidWeightedLinearUnits2017} nonlinearity and a skip connection~\cite{heDeepResidualLearning2015}. We also apply layer normalization \cite{baLayerNormalization2016} across the lattice dimensions, as illustrated in Fig.~\ref{fig:nn-diagram}. Mathematically, the output $\hh'$ of each block is
\begin{equation}
\label{eq:res-block-def}
    \hh' = \hh + \textup{Conv}_{\downarrow} \left( \textup{SiLU} \left( \textup{Conv}_{\uparrow} \left( \textup{Norm} \left( \hh \right) \right)  \right) \right)
\end{equation}
for a given input $\hh$. Convolutions in Eq.~\eqref{eq:res-block-def} are defined as $ \textup{Conv} (\hh) ^k = b^k + \sum _l w^{k l} \ast \hh ^l$,
with $\ast$ being the periodic (antiperiodic) convolution operation in the case of periodic (antiperiodic) boundary conditions on $\xx$ or the zero-padded convolution in the case of open boundary conditions. By choosing the appropriate convolution, we keep each of the residual blocks and the overall model translationally invariant. The first convolution, $\textup{Conv}_{\uparrow}$, increases the number of features to $\alpha \, d$ before the second convolution $\textup{Conv}_{\downarrow}$ decreases it back to $d$.

Layer normalization is applied across the lattice dimension for each feature or channel independently as
\begin{equation}
    \left( \textup{Norm} (\hh) \right) ^k = \frac{\hh ^k - \expect [\hh ^k] }{ \sqrt{ \Var [\hh ^k] + \epsilon}} \times \gamma ^k + \beta ^k \; ,
\end{equation}
where $\expect [\cdot]$ and $\Var [\cdot]$ are the empirical mean and variance, respectively, and the scale $\gamma$ and the shift $\beta$ are trainable parameters, with $\epsilon$ fixed to a small constant for numerical stability in cases of vanishing variance.

We produce the final hidden spin values as
\begin{equation}
\label{eq:hidden-spin-proj}
    \vb{y} = \textup{SoftSign} \left( \textup{Norm} \left( \hh \right) \right)
\end{equation}
from the output residual blocks $\hh$. We emphasize that each component of $\vb{y}$ is restricted to $-1 < y_i < 1$ by this operation. The \textup{SiLU}~\cite{elfwingSigmoidWeightedLinearUnits2017} and \textup{SoftSign} nonlinearities in Eqs.~\eqref{eq:res-block-def}~and~\eqref{eq:hidden-spin-proj} are defined as
\begin{equation}
    \textup{SiLU}(x) = \frac{x}{1 + e^{-x}}
    \quad \text{and} \quad
    \textup{SoftSign}(x) = \frac{x}{1 + |x|} \; ,
\end{equation}
and are always applied elementwise. Numerical values of all hyperparameters can be found in the Supplemental Material~\cite{supplement}.

\bibliography{references-matija, references-alev}

\end{document}


\title{
    Adiabatic transport of neural network quantum states\\
    Supplementary material
}

\author{Matija Medvidović}
\email{mmedvidovic@ethz.ch}
\affiliation{Institute for Theoretical Physics, ETH Zürich, CH-8093 Zürich, Switzerland}

\author{Alev Orfi}
\affiliation{Center for Computational Quantum Physics, Flatiron Institute, 162 5th Avenue, New York, NY 10010, USA}
\affiliation{Center for Quantum Phenomena, Department of Physics, New York University, 726 Broadway, New York, New York 10003, USA}

\author{Juan Carrasquilla}
\affiliation{Institute for Theoretical Physics, ETH Zürich, CH-8093 Zürich, Switzerland}

\author{Dries Sels}
\affiliation{Center for Computational Quantum Physics, Flatiron Institute, 162 5th Avenue, New York, NY 10010, USA}
\affiliation{Center for Quantum Phenomena, Department of Physics, New York University, 726 Broadway, New York, New York 10003, USA}

\date{\today}

\maketitle
\onecolumngrid
\vspace{-10ex}

\section{Neural-network quantum state optimization}

In this short appendix, we repeat the details from Refs.~\cite{medvidovicVariationalQuantumDynamics2023, schmittQuantumManyBodyDynamics2020} around the linear solver used to solve the parameter update equation $\GG \delta \theta = - f$. We begin by performing a singular value decomposition (SVD) of the matrix $A$, representing either $J^\top P$ or $P J^\top$, depending on the version of the parameter update rule presented in the main text. We seek to approximate $A^{-1}$. By employing standard linear algebra routines, we get orthogonal matrices $U$ and $V$ and a diagonal matrix $\Sigma = \diag (\sigma _1, \sigma _2, \ldots )$ with nonnegative real entries satisfying $A = U \Sigma V^\top$. If all singular values $\sigma _i$ are positive and sufficiently removed from zero (depending on the floating point data type and machine precision), then the inverse exists and can be computed as $A^{-1} = V \Sigma ^{-1} U^\top$.

\begin{figure*}[h]
    \centering
    \begin{overpic}[width=1\linewidth]{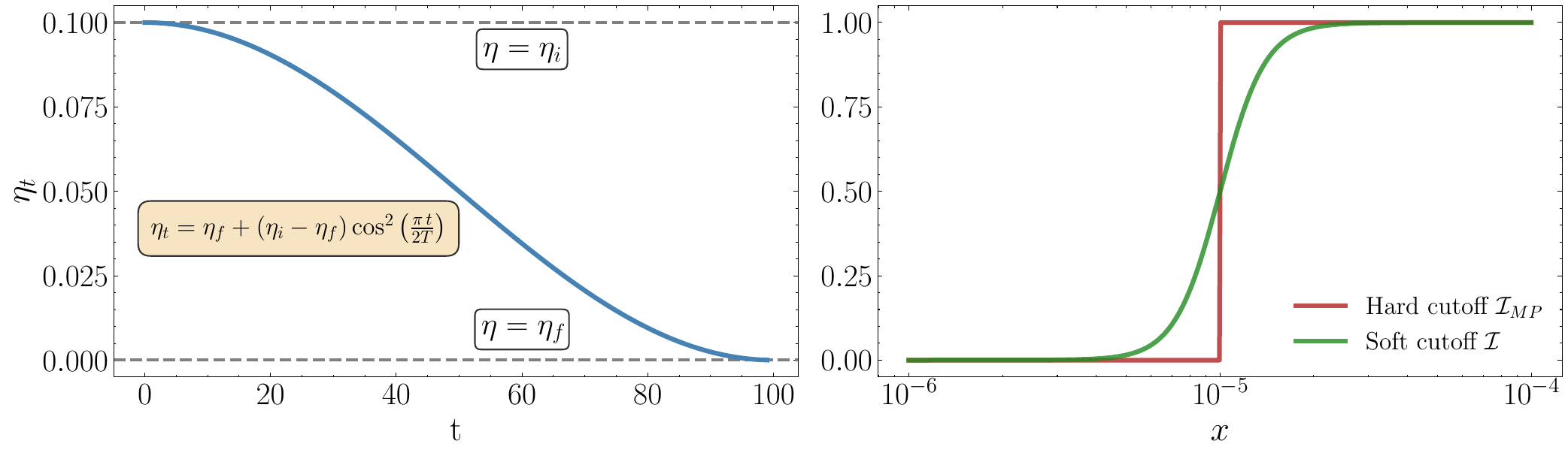}
        \put(8, 25){\fontsize{10}{12}\selectfont\textbf{(a)}}   
        \put(57, 26){\fontsize{10}{12}\selectfont\textbf{(b)}}
    \end{overpic}
    \caption{
    The \textit{dampening} (learning rate) schedule, showing the cosine decay in Eq.~\ref{eq:lr-schedule} (a) and the comparison between the Moore-Penrose weighting of inverse singular values and the \textit{soft} pseudoinverse used in our simulations (b).
    }
    \label{fig:dampening-and-pseudoinverse}
\end{figure*}

However, when some of the singular values are nearly or identically zero, matrix inversion can only be performed in the subspace spanned by the singular vectors corresponding to (sufficiently) nonzero singular values. Instead of truncating all singular values below some cutoff $\epsilon$, we opt to perform a \textit{soft} inverse $\mathcal{I}$. We construct $A^{-1} \approx V \, \mathcal{I} \left( \Sigma \right) \, U^\top$, where
\begin{equation}
\label{eq:pseudoinverse}
    \mathcal{I} (x) = \frac{\nicefrac{1}{x}}{1 + \left(\frac{\epsilon}{x} \right)^6} \; .
\end{equation}
The \textit{soft} pseudoinverse defined in Eq.~\ref{eq:pseudoinverse} is contrasted with the standard Moore-Penrose pseudoinverse in Fig.~\ref{fig:dampening-and-pseudoinverse}, which is given by the step function $\mathcal{I} _{\text{MP}} (x) = \flatfrac{\theta (x - \epsilon)}{x}$. Given a cutoff $\epsilon$, some singular values are exactly inverted, while others are set to zero.

\begin{figure}[h]
    \centering
    \begin{overpic}[width=\linewidth]{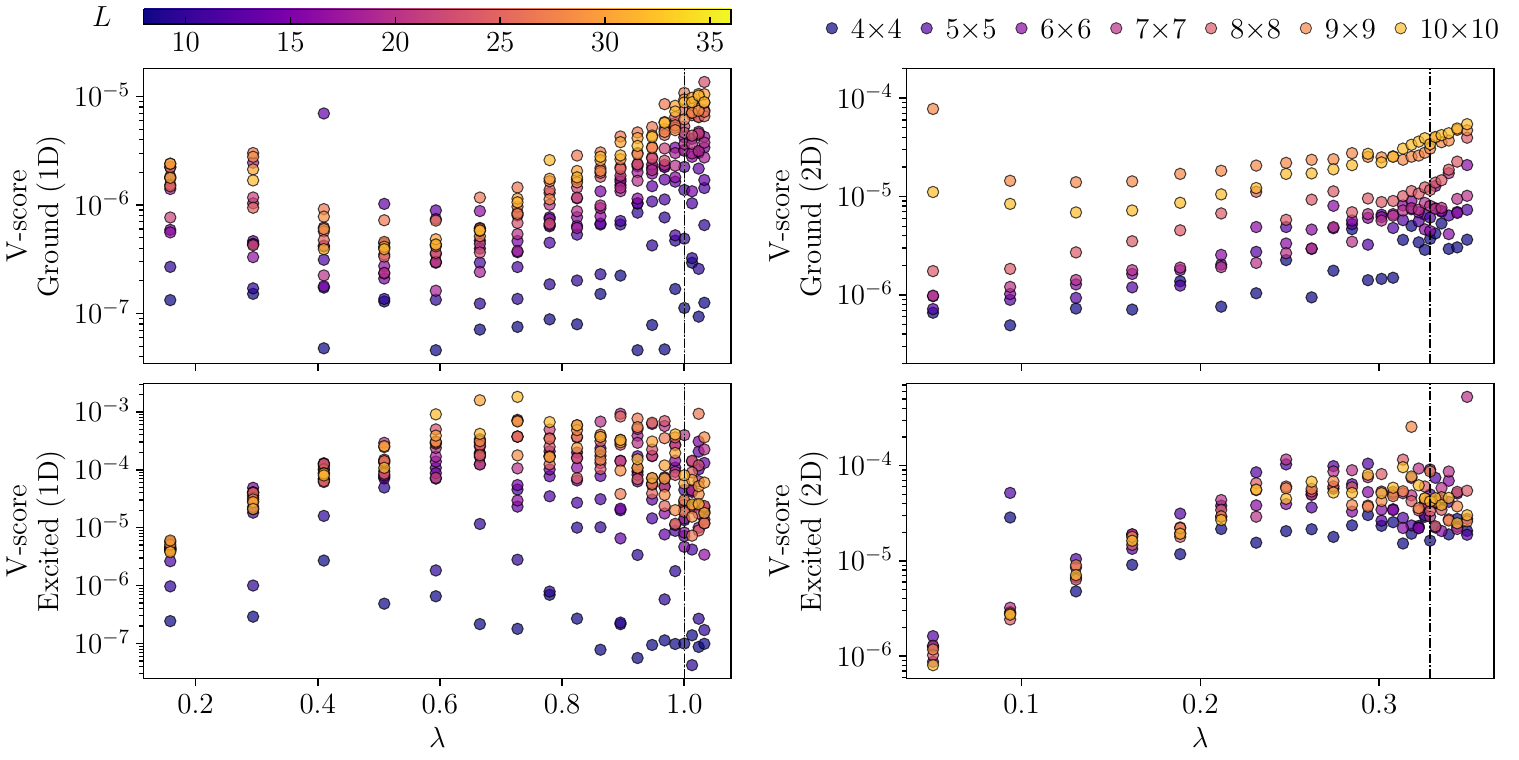}
        \put(10.5, 43){\fontsize{10}{12}\selectfont\textbf{(a)}}   
        \put(10.5, 22){\fontsize{10}{12}\selectfont\textbf{(b)}}
        \put(63, 43){\fontsize{10}{12}\selectfont\textbf{(c)}}
        \put(61, 22){\fontsize{10}{12}\selectfont\textbf{(d)}}
    \end{overpic}
    \caption{
        V-scores of transported ground and excited states of the 1D [(a),(b)] and 2D [(c),(d)] TFIM corresponding to the data in presented in the main text.
    }
    \label{fig:vscore}
\end{figure}

For each change in the Hamiltonian parameter $\lambda$, a maximum of $M$ inverse power iterations is performed, with $\eta$ annealed from $\eta_i$ to $\eta_f$ following a cosine schedule,
\begin{equation}
\label{eq:lr-schedule}
    \eta _t = \eta _f + (\eta _i - \eta _f) \cos ^2 \left( \frac{\pi \, t}{2 T} \right) \; .
\end{equation}
The IPI process was terminated prior to reaching the maximum of $M$ iterations if the energy variance was below a cutoff (1D: $5\times10^{-7}$ and 2D: $5\times10^{-5}$), indicating sufficient convergence to an eigenvector. Final V-scores of the transported states used for the critical exponent analysis are shown in Fig.~\ref{fig:vscore}. The transport hyperparameters, listed in Table~\ref{tab:hyperparams}, were selected with minimal optimization, yet they yielded low energy variances across the considered range of $\lambda$. This suggests that further reductions in computational resources may be possible. It was found that scaling the spin embedding dimension and the number of samples with system size was beneficial in one dimension. This is likely due to the nonlocal nature of domain walls in the excited state, as such scaling was not necessary in two dimensions.

\begin{table*}[h]
\centering
\begin{tabular}{c|c|c|c|c|c}
    \textbf{Symbol} & \textbf{Name} & \textbf{Value 1D} & \textbf{Value 2D} & \textbf{Domain} & \textbf{Description} \\
    \hline \hline
    $B$ & \thead{Number of\\residual blocks} & 2 & 2& $\mathbbm{N}$ & Residual block count \\
    \hline
    $d$ & \thead{Spin embedding\\dimension} & $N/2$ & 8& $\mathbbm{N}$ & \thead{Dimension of internal spin \\ representations within the model} \\
    \hline
    $\eta$ & \thead{Dampening\\(Learning rate)} & \thead{Scheduled \\ $0.02 \rightarrow 0.01$} & \thead{Scheduled \\ $0.02 \rightarrow 0.0005$} & $\mathbbm{R}_+$ & IPI update multiplier for numerical stability \\
    \hline
    $M$ & \thead{Number of inverse\\power iterations} & 80 & 100 & $\mathbbm{N}$ & Number of IPI for each new state\\
    \hline
    $\alpha$ & Enhancement & 2 & 2 & $\mathbbm{R} _+$ & \thead{Relative increase in the number of features\\in the middle \texttt{Conv} layer in each residual block.}\\
    \hline
    $N_s$ & Number of samples & $N$&32 & $\mathbbm{N}$ & \thead{The number of Monte Carlo\\samples per chain }\\
    \hline
    $N_c$ & Number of chains & 12&32 & $\mathbbm{N}$ & \thead{The number of Monte Carlo chains}\\
    \hline
    $S$ & Number of steps & 20  & 25 &$\mathbbm{N}$  & \thead{The number of steps of $\lambda$ values\\ between $\lambda_0$ and $\lambda_f$}
\end{tabular}
\caption{
    The list of relevant hyperparameter choices used in this work.
}
\label{tab:hyperparams}
\end{table*}

\section{Data analysis}

\begin{figure*}[h]
    \begin{overpic}[width=1\linewidth]{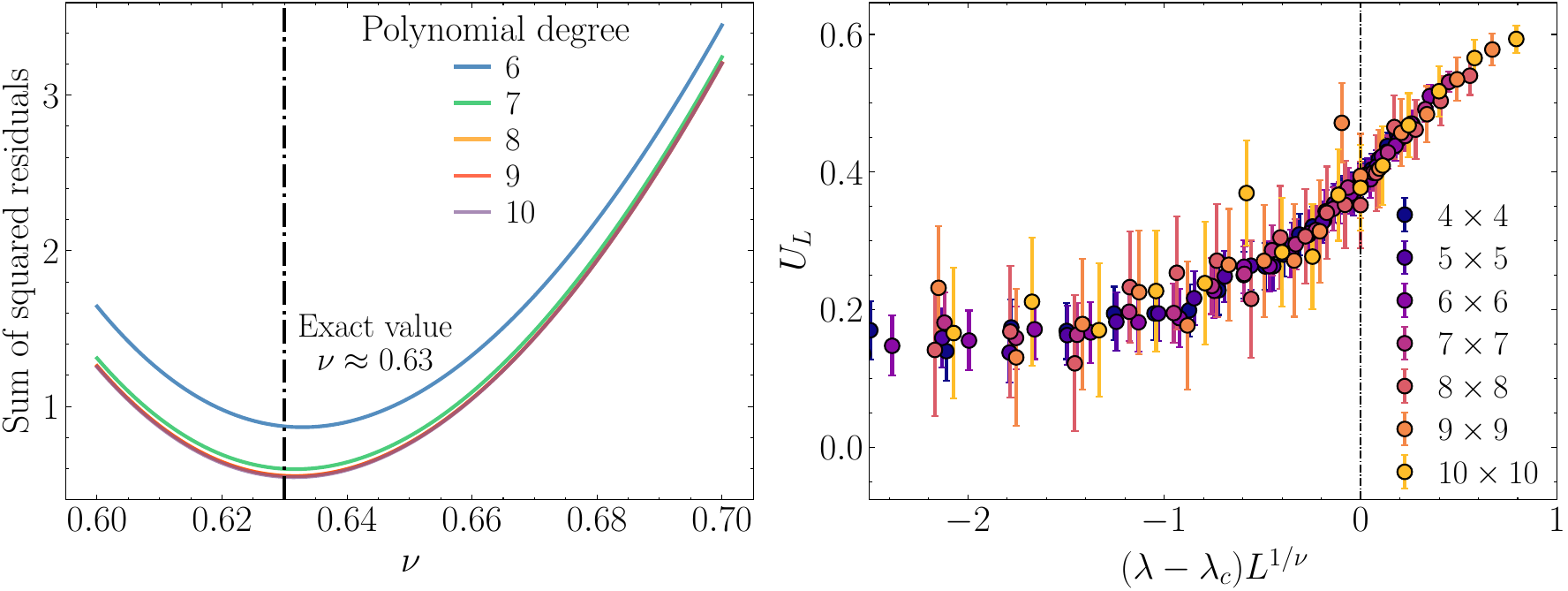}
        \put(6, 35){\fontsize{10}{12}\selectfont\textbf{(a)}}   
        \put(57, 35){\fontsize{10}{12}\selectfont\textbf{(b)}}
    \end{overpic}
    \caption{(a) Sum of squared residuals for data collapse used to estimate ${\nu}$. (b) Binder cumulant $U_L$ collapsed using the extracted critical exponents, confirming the critical scaling.
    }
    \label{fig:polynomial-fit}
\end{figure*}

Critical exponents in tabulated in the main text have been calculated by statistical analysis of the adiabatically transported energy levels $E_{n,\lambda}$. In this appendix, we briefly detail the steps we take to extract numerical values and error bars. We use the known critical scaling relation $\Delta \sim L ^{-z}$ at the critical point to recover an estimate $\Hat{z}$ using a least-squares linear fit in the log space. On the other hand, $\nu$ is determined by a method of sequential polynomial fits to the scaling relation given in the main text. Given that $F$ is a universal function (independent of the system size $L$), we expand it as a degree $D$ polynomial with universal coefficients $c_k$
\begin{equation}
    \Delta \; L^z = F \left( \left( \lambda - \lambda_c \right) L^{\nicefrac{1}{\nu}} \right)
    \qquad \longrightarrow \qquad
    F(x) = \sum _{k=0} ^{D} \, c_k \, x^k \; .
\end{equation}
For a chosen $\nu$, we perform a least-squares fit to estimate the coefficients $c_k$ and the corresponding sum of squared residuals $\text{SSR} (\nu)$. Temporarily restoring all of the suppressed notation for clarity, we have
\begin{equation}
    \text{SSR} (\nu) = \min _{c_k} \left\{ \sum _{\lambda, L} \left[ \Delta _L (\lambda) \; L^{\Hat{z}} - \sum _{k=0} ^{D} \, c_k \, ( \lambda - \lambda_c) ^k L^{\nicefrac{k}{\nu}} \right] ^2 \right\} \; .
\end{equation}
Our estimate $\Hat{\nu}$ is then simply $\Hat{\nu} = \argmin _{\nu} \; \text{SSR}(\nu)$. Being a measure of the \textit{goodness-of-fit}, minimizing SSR is equivalent to minimizing the overall vertical spread of data points with respect to the best-fit polynomial, requiring that the function $F$ be as close to well defined or single-valued as possible. From Fig.~\ref{fig:polynomial-fit} (a), we see that it is sufficient to choose polynomials of degree $D \gtrsim 8$. The final value of $\Hat{\nu}$ was obtained by numerical minimization of the SSR.

In traditional quantum Monte Carlo calculations, where excited state information is unavailable, critical exponents are commonly extracted through the Binder cumulant defined as
\begin{equation}
    U_L = 1 - \frac{\expect [m (\xx)^4]}{3 \expect [m (\xx) ^2]}
    \qquad \text{where} \qquad
    m (\xx) = \frac{1}{N} \sum _{\sigma \in \xx} \sigma
\end{equation}
is the magnetization per spin. The Binder cumulant is used because it has a vanishing dynamical critical exponent $z_U = 0$, leading to the scaling relation
\begin{equation}
    U_L (\lambda ) = F_U \left( \left( \lambda - \lambda_c \right) L^{\nicefrac{1}{\nu}} \right),
\end{equation}
which we can use to validate our found exponents. Fig.~\ref{fig:polynomial-fit}(b) shows that the exponents obtained by fitting the gap scaling relation also collapse the Binder cumulant data.

Estimated exponents are used to collapse the fidelity susceptibility, accounting for finite-size shifts of the critical point. In one dimension, where the exact fidelity is known (see next section), the shift is given by $\lambda^* = \lambda_c - \frac{\pi^2}{2N^2}$. Whereas in two dimensions, the following shift is considered $\lambda^* = \lambda_c + b L^{-1/\nu}$. The coefficient $b$ is determined by fitting the fidelity susceptibility to a Lorentzian and minimizing the peak distance from the known critical point.

\section{Ising chain fidelity susceptibility}
\label{appendix:fid_susc}

The fidelity susceptibility of the one-dimensional Ising chain can be obtained from the exact solution of the model and was used as a point of comparison in the main text. The Ising Hamiltonian can be recast in terms of fermionic operators using the Jordan–Wigner transformation \cite{sachdevQuantumPhaseTransitions2011}, 
\begin{equation}
     H = -\lambda\sum_i^{N-1}(a_{i}^\dagger-a_i)(a_{i+1}+a_{i+1}^\dagger) -\lambda e^{i\pi \hat{N}} (a_{N}^\dagger-a_N)(a_{1}+a_{1}^\dagger)+ \sum_i^{N-1} (2a_i^\dagger a_i -1).
\end{equation}
The original $\mathbb{Z}_2$ symmetry of the spin system is now mapped to the fermionic parity $p = \frac{1}{2}(1-e^{i\pi \hat{N}})$, which allows the two parity sectors to be analyzed independently. Within each parity sector, the Hamiltonian takes the form,
\begin{equation}
     H_p = - \lambda\sum_i^{N}(a_{i}^\dagger-a_i)(a_{i+1}+a_{i+1}^\dagger) + \sum_i^N (2a_i^\dagger a_i -1)
\end{equation}
with the boundary conditions $a_{N+1} = (-1)^{p+1}a_{1}$. Transforming into momentum space, $a_k = \frac{1}{\sqrt{N}}\sum_{j=1}^N e^{-ikj}a_j$, gives the following form of the Hamiltonian,
\begin{equation}
    H_p = \sum_{K_p}\left[\left(1-\lambda\cos(k)\right)\left(a_k^\dagger a_k-a_{-k}a_{-k}^\dagger\right) +i\lambda\sin(k)\left(a_{-k} a_{k}-a_k^\dagger a_{-k}^\dagger\right)\right]
\end{equation}
where the allowed momenta depend on the parity sector,
\begin{equation}\label{eq:kvalues}
K_p = \bigg\{k= \pm\frac{2\pi}{N} \times \begin{cases}
(l-1/2) \quad \text{with } l= 1,2,...,N/2 & p=0\\
l \quad \text{with } l= 1,2,...,N/2-1 & p=1
\end{cases}
\end{equation}
The momentum values group into pairs $(k,-k)$ enabling the Hamiltonian to be written as the sum over positive $k$ values, $K_p^+$,
\begin{equation}
    H_{0} = \sum_{K^+_0} H_k, \quad\quad\quad H_1 = \sum_{K^+_1} H_k + H_{k=0,\pi}. 
\end{equation}
with the $k=0$ and $k=\pi$ contributions in the $p=1$ sector are, 
\begin{equation}
    H_{k=0,\pi} =2(\hat{n}_\pi-\hat{n}_0)+2h(\hat{n}_0+\hat{n}_\pi-1).
\end{equation}
Each block Hamiltonian can now be written as
\begin{equation}
     H_k = 2\begin{pmatrix}
    a_k^\dagger & a_{-k}
    \end{pmatrix}
    \mathbf{H}_k
    \begin{pmatrix}
    a_k \\ a_{-k}^\dagger
    \end{pmatrix}
\end{equation}
where, 
\begin{equation}
    \mathbf{H}_k = \begin{pmatrix}
    1-\lambda\cos(k) & -i\lambda\sin(k) \\ 
    i\lambda\sin(k) & -1+\lambda\cos(k)
    \end{pmatrix} = A_k(\lambda)\sigma_k^z+B_k(\lambda)\sigma_k^y.
\end{equation}
Diagonalization is achieved via a Bogoliubov rotation,
\begin{equation}
    \gamma_k = u_k a_k -iv_k a_{-k}^\dagger
\end{equation}
with $u_k = \cos(\theta_k/2)$ and $v_k = \sin(\theta_k/2)$, with angle $\theta_k$ defined as, 
\begin{equation}
        \cos(\theta_k) = \frac{1 - \lambda\cos(k)}{\sqrt{(1 - \lambda\cos(k))^2+\lambda^2\sin^2(k)}},\;\;
        \sin(\theta_k) = \frac{\lambda\sin(k)}{\sqrt{(1 - \lambda\cos(k))^2+\lambda^2\sin^2(k)}}.
\end{equation}
In terms of these Bogoliubov fermions, the Hamiltonian takes the diagonal form, 
\begin{equation}
    H_k =  2\varepsilon_k(\gamma_k^\dagger\gamma_k-1/2) \; ,
\end{equation}
with dispersion relation $\varepsilon_k = \sqrt{ ( 1 - \lambda \cos(k) ) ^2 + \lambda^2 \sin^2 (k)}$. This fermionic representation allows the ground-state fidelity susceptibility to be computed in two complementary ways. One approach is to evaluate the fidelity directly between ground states at couplings differing by an infinitesimal $\epsilon$ in $\lambda$. Alternatively, it can be written in terms of the adiabatic gauge potential (AGP) $\mathcal{A}_\lambda$,
\begin{equation}
    \chi_0(\lambda) = \bra{\Psi_0(\lambda)} \mathcal{A}_\lambda^2 \ket{\Psi_0(\lambda)},
\end{equation}
where $\ket{\Psi_0(\lambda)}$ is the ground state of the system \cite{kolodrubetzGeometryNonadiabaticResponse2017}. The ground state of the system, known as the Bogoliubov vacuum, is the state that annihilates $\gamma_k$ for all $k$. For a single momentum mode, this state is
\begin{equation}
    \ket{\Psi_0^k(\lambda)} = \cos\left(\frac{\theta_k(\lambda)}{2}\right)\ket{0_k} + i\sin\left(\frac{\theta_k(\lambda)}{2}\right)\ket{1_k}.
\end{equation}
The AGP can be found variationally in each of the non-interacting $k$-blocks independently \cite{kolodrubetzGeometryNonadiabaticResponse2017}. Consider the general variational form, 
\begin{equation}
    \mathcal{A}_\lambda(k) = \frac{1}{2}\left(\alpha_x(k)\sigma_k^x+\alpha_y(k)\sigma_k^y+\alpha_z(k)\sigma_k^z\right),
\end{equation}
where the $\alpha_x$,$\alpha_y$, and $\alpha_z$ corresponding to the AGP minimize the action~\cite{kolodrubetzGeometryNonadiabaticResponse2017}, 
\begin{equation}
    S = \Tr\left[G^2_\lambda(\mathcal{A}_\lambda(k) )\right], \;\; G_\lambda(\mathcal{A}_\lambda(k) ) = \partial_\lambda\mathbf{H}_k  + i\left[\mathcal{A}_\lambda(k), \mathbf{H}_k \right].
\end{equation}
For a momentum mode, this gives
\begin{equation}
    \begin{split}
        G_\lambda(\mathcal{A}_\lambda(k) ) &= -\cos(k)\sigma_k^z+\sin(k)\sigma_k^y +\left((\alpha_zB_k-\alpha_yA_k)\sigma_k^x+\alpha_xA_k\sigma_k^y-\alpha_xB_k\sigma_k^z\right),\\
        &= (\alpha_zB_k-\alpha_yA_k)\sigma_k^x +(\alpha_xA_k+\sin(k))\sigma_k^y +(-\alpha_xB_k-\cos(k))\sigma_k^z \; .
    \end{split}
\end{equation}
Minimizing the action with respect to $\alpha_x$, $\alpha_y$, and $\alpha_z$ gives the following form of the adiabatic gauge potential for a $k$ block,
\begin{equation}
    \mathcal{A}_\lambda(k) = -  \frac{A_k\sin(k)+B_k\cos(k) }{2(A_k^2+B_k^2)}\sigma_k^x = -  \frac{\sin(k) }{2\varepsilon_k^2}\sigma_k^x.
\end{equation}
Since both the ground state and $\mathcal{A}_\lambda$ decompose into independent $k$-blocks, the fidelity susceptibility takes the form
\begin{equation}
    \chi_0(\lambda) = \sum_k \bra{\psi_0^k(\lambda)} \mathcal{A}_\lambda(k)^2 \ket{\psi^k_0(\lambda)}
    = \sum_k \frac{\sin^2(k) }{4\varepsilon_k^4}.
\end{equation}
From the ground state description, the infinitesimal fidelity can also be calculated directly,
\begin{equation}
    F_0(\lambda,\lambda+\epsilon) = \prod_k |\braket{\Psi_0^k(\lambda)}{\Psi_0^k(\lambda+\epsilon)}|^2 = \prod_k \cos^2\left(\frac{1}{2}(\theta_k(\lambda)-\theta_k(\lambda+\epsilon)\right).
\end{equation}
This can additionally be used to get the following fidelity susceptibility that matches the result above:
\begin{equation}
    \begin{split}
    \chi_0 &=  \lim_{\epsilon \to 0} \frac{\partial^2}{\partial \epsilon^2} \left( -\ln F_0(\lambda,\lambda+\epsilon) \right),\\
    &=  \lim_{\epsilon \to 0} \frac{\partial^2}{\partial \epsilon^2} \left(-\sum_k \ln \left(\cos^2\left(\frac{1}{2}(\theta_k(\lambda)-\theta_k(\lambda+\epsilon)\right)\right) \right),\\
    &=  \lim_{\epsilon \to 0} \frac{\partial^2}{\partial \epsilon^2}\left(-\sum_k \ln \left(\cos^2\left(\frac{1}{2}\epsilon\frac{\partial\theta_k}{\partial\lambda}\right)\right) \right),\\
    &=  \lim_{\epsilon \to 0} \frac{1}{4}\sum_k\frac{\partial^2}{\partial \epsilon^2} \left[\epsilon^2 \left(\frac{\partial\theta_k}{\partial\lambda}\right)^2\right],\\
    & = \sum_k \frac{\sin^2(k)}{4\varepsilon_k^4}.
    \end{split}
\end{equation}

\bibliography{references-matija, references-alev}